\documentclass[12pt,a4paper]{amsart}

\usepackage{amsfonts}
\usepackage{amssymb}
\usepackage{dsfont}
\usepackage{enumerate}
\usepackage{xspace}
\usepackage[all]{xy}

\def \J{\mathcal{J}}
\def \A{\mathcal{A}}
\def \L{\mathcal{L}}
\def \S{\mathcal{S}}
\def \t{\widetilde}
\def \F{\mathcal{F}}
\def \M{\mathcal{M}}
\def \D{\mathcal{D}}
\def \O{\mathcal{O}}

\def \N{\mathcal N_{\mathcal {J}}}
\def \Nc{\mathcal N_{\mathcal {J}}^{\mathbb{C}}}
\def \C{\mathbb{C}}
\def \R{\mathbb{R}}
\def \Hw{\mathcal H_{\t \omega}}
\newcommand{\LJB}{\mbox{LJB--algebra}\xspace}
\newcommand{\LJBs}{\mbox{LJB--algebras}\xspace}
\newcommand{\CA}{C^*\mbox{--algebra}\xspace}

\newtheorem{theorem}{Theorem}
\newtheorem{corollary}{Corollary}

\newtheorem{definition}{Definition}
\newtheorem{lemma}{Lemma}

\newcommand{\bigslant}[2]{{\left.\raisebox{.2em}{$#1$}\!\middle/\!\raisebox{-.2em}{$#2$}\right.}}

\begin{document}

\title{Reduction of Lie--Jordan Banach algebras and quantum states}

\author{F. Falceto$^a$, L. Ferro$^b$, A. Ibort$^b$ and  G. Marmo$^{b,c}$}

\address{$^a$ Departamento de F\'{\i}sica Te\'orica. Universidad de Zaragoza\\ Plaza San Francisco s/n, 50009 Zaragoza, Spain}

\address{$^b$ Departamento de Matem\'aticas, Universidad Carlos III de Madrid\\ Avda. de la Universidad 30, 28911 Legan\'es, Madrid, Spain}

\address{$^c$ Dipartimento di Scienze Fisiche, INFN--Sezione di Napoli, Universit\`a di Napoli ``Federico II'', Via Cintia Edificio 6, I--80126 Napoli, Italy}

\email{falceto@unizar.es ,lferro@math.uc3m.es, albertoi@math.uc3m.es, marmo@na.infn.it}

\begin{abstract} A theory of reduction of Lie--Jordan Banach algebras with respect to either a Jordan ideal or a Lie-Jordan subalgebra is presented.   This theory is compared with the standard reduction of $\CA$s of observables of a quantum system in the presence of quantum constraints.  It is shown that the later corresponds to the particular instance of the reduction of Lie--Jordan Banach algebras with respect to a Lie--Jordan subalgebra as described in this paper.  The space of states of the reduced Lie--Jordan Banach algebras is described in terms of equivalence classes of extensions to the full algebra and their GNS representations are characterized in the same way.   A few simple examples are discussed that illustrates some of the main results.

\end{abstract}

\maketitle


\section{Introduction}  

In this paper we present a theory of reduction of Lie--Jordan algebras that can be used as an alternative to deal with symmetries and local constraints in quantum physics and quantum field theories.    R. Haag's algebraic approach to quantum systems \cite{Ha96} has had a profound influence in both the foundations and applications of quantum physics. The background for that approach is to consider a quantum system described by a $\CA$ $\A$ of observables, its quantum states $\omega$ are normalized positive complex functionals on it. The state space $\mathcal S$ becomes a convex $\mathrm{w}^*$--compact topological space and pure states are its extremal points. Moreover, given a state $\omega$, the GNS construction allows to represent the $\CA$ $\A$ on a cyclic complex Hilbert space $\mathcal H_\omega$.    

A geometrical approach to Quantum Mechanics \cite{Carinena:2007ws}, \cite{Chruscinski:2008px} has been developed in the last twenty years (see also Cirelli and Lanzavecchia \cite{Ci84} and Ashtekar and Schilling  \cite{As99}). After this, it has become obvious that the geometrical picture of Quantum Mechanics could be useful to gain insight into such properties of quantum systems like integrability \cite{ClementeGallardo:2008wa}, the intrinsic nature of different measures of entanglement \cite{Grabowski:2000zk},\cite{Grabowski:2005my}, etc. (see for instance \cite{GMS} and references therein).

The geometrical description of dynamical systems provides a natural setting to describe systems with symmetry and/or constraints.  If the system carries a symplectic or Poisson structure several procedures were introduced along the years to cope with the different cases, like Marsden-Weinstein reduction, symplectic reduction, Poisson reduction, reduction of contact structures, etc.   Again, it was soon realized that the algebraic approach to reduction provided a convenient setting to deal with reduction of systems \cite{Ib97} \cite{Gr94}.

Whenever constraints are imposed on a quantum system or symmetries are present, both dynamical or gauge, some reduction on the state space must be considered either because not all states are physical and/or because families of states are equivalent.  In the standard approach to Quantum Mechanics, constraints are imposed on the system by selecting subspaces determined by the quantum operators corresponding to the constraints of the theory, Dirac states, and equivalence of quantum states was dealt with by using the representation theory of the corresponding group of symmetries.  Many difficulties and ambiguities emerge when implementing this analysis as different classes of singular Lagrangian systems are considered or other singularities were found (quantization of singular level sets of momentum maps for instance).

Taking as a departing point for the analysis the algebraic approach to Quantum Mechanics and Quantum Field Theory the problem of reduction of the quantum system becomes the problem of reducing the $\CA$ of observables. Such programme was succesfully developed for some gauge theories and the proposed reduction of the algebra of observables was called T--reduction (see Grundling et al \cite{Grundling:1984sq} and references therein) and more generally for quantum local constraints it was discussed at \cite{Grundling:1998zn}.   There is however a point that should be raised here, and it is that the state space $\mathcal S$ of the quantum system does not determine univocally the $\CA$ structure of the algebra of observables but only its Jordan Banach real algebra \cite{Ne34} \cite{Se47}. In fact the real (or self-adjoint) part of a $\CA$ $\mathcal A$, because of a theorem by Kadison is isometrically isomorphic to the space of all bounded affine functions on the state space \cite{Ka51}.   Then the question of when a given Jordan Banach algebra is the real part of a $\CA$ raises.   A. Connes on one side \cite{Co74} and Alfsen and Shultz on the other \cite{Al98}, gave different answers to this question.    In the present context the characterization obtained by Alfsen and Schultz in terms of the so--called dynamical correspondence on a Jordan Banach algebra amounts to state that the relevant structure to discuss properties of the state space of a quantum system is that of Lie--Jordan Banach algebra \cite{Em84} \cite{La98}. In fact the topological properties of the state space are completely captured by the Jordan Banach--algebra structure and the Lie algebra structure allows to construct the $\CA$ setting for them, their GNS representations, etc.\\

Thus, as stated above, we will address the problem of the reduction of Quantum systems as the reduction of Lie--Jordan Banach algebras. Inspired by the general reduction theory of Jordan algebras (briefly reviewed in section \ref{LJ red}) and combining it with the theory of reduction of (noncommutative) Poisson algebras, we will obtain a general reduction theorem for Lie--Jordan Banach algebras with respect to Jordan ideals, Sect. \ref{LJ red}, and with respect to Lie--Jordan subalgebras, Sect. \ref{redsub}. These results will be compared to the theory of reduction of $\CA$ with respect to quantum constraints that will be shown to be a particular instance of the theory discussed in this paper, Sect. \ref{constraints}.   An explicit description of the states of the reduced Lie--Jordan Banach algebras will be obtained and the corresponding GNS representations will be discussed in Sects. \ref{reduced_states} and \ref{reduced_gns}.

\section{Quantum states and Lie--Jordan Banach algebras}\label{second}

Quantum states are positive linear functionals $\omega$ on a $\CA$ $\A$ of observables, thus $\omega(a^*a) \geq 0 \ \forall a \in \A$. We will assume in what follows that the $\CA$ is unital and states are normalized, i.e. $\omega(\mathds{1}) = 1$, where $\mathds{1}$ denotes the unit element of $\A$. We will denote by $\A_{\mathrm{sa}}$ the real (or self-adjoint) part of $\A$, i.e. 
$$\A_{\mathrm{sa}} = \{ a \in \A\ |\ a^* = a\},$$ and we will denote by $\| \cdot \|$ the induced norm. The self-adjoint part of $\A$ inherits the structure of a real Lie--Jordan algebra. It is a Jordan algebra with the non-associative commutative Jordan product
\begin{equation}\label{jordan}
a \circ b = \frac{1}{2} (ab + ba),\quad \forall a, b \in \A_{\mathrm{sa}},
 \end{equation}
and by setting $a \circ a = a^2$, the Jordan product $\circ$ satisfies
\begin{equation}\label{jassociativity}
(a^2 \circ b) \circ a = a^2 \circ (b \circ a) 
\end{equation}
which is the usual replacement for associativity for Jordan algebras.\\
The Lie product is obtained by taking a scaled commutator:
\begin{equation}\label{lie}
 [a,b]_\lambda = i\lambda (ab - ba),\quad \forall a, b \in \A_{\mathrm{sa}},
 \end{equation}
with $\lambda \in \mathbb{R}$. Notice that $[a,b]_\lambda \in \A_{\mathrm{sa}}\ \forall a, b \in \A_{\mathrm{sa}}$.\\ The skew-symmetric bilinear form $[ \cdot, \cdot]_\lambda$ equips $\A_{\mathrm{sa}}$ with a Lie algebra structure since it verifies the Jacobi identity
\begin{equation}\label{jacobi}
 [[a,b]_\lambda,c]_\lambda + [[c,a]_\lambda,b]_\lambda + [[b,c]_\lambda,a]_\lambda = 0.
\end{equation}
These two operations are compatible in the sense that the following Leibniz identity is satisfied:
\begin{equation}\label{leibniz}
 [a,b\circ c]_\lambda = [a,b]_\lambda\circ c + b\circ [a,c]_\lambda,
\end{equation}
or, in other words, the linear map $D_a(\cdot) \equiv [a,\cdot\ ]_\lambda$ is a derivation of the Jordan product $\circ$.

A vector space
$V$ with two operations $\circ$ and $[\cdot ,\cdot ]$ satisfying the properties of the operations
above, eqs. (\ref{jordan})-(\ref{leibniz}), is called a (pre-) Lie--Jordan algebra.  The complete definition
of a Lie--Jordan algebra requires that the associator of the structure product is
related to the Lie bracket by:
\begin{equation}\label{associator}
 (a\circ b)\circ c - a \circ (b \circ c) = \kappa [b,[c,a]_\lambda]_\lambda ,
\end{equation}
$\kappa$ being a positive real number.\\
When using the Lie bracket (\ref{lie}), we obtain:
$$ k \lambda^2 = 1/ 4 .$$
A Lie--Jordan algebra will be called unital if it contains the unit element with respect to the Jordan product.\\

Moreover if a Lie--Jordan algebra $\L$ carries a complete norm $\| \cdot \|$ such that $\forall a,b \in \L$:
\begin{enumerate}[i)]
\item $\| a \circ b\| \leq \|a\|\ \|b\|$
 \item $\| a \circ a \| = \| a\|^2$
\item $\|a \circ a\| \leq \|a \circ a + b \circ b\|$ for all $a,b \in \L$,
\end{enumerate}
and $[\cdot,\cdot]_\lambda$ is continous, then $\L$ is a Lie--Jordan Banach algebra (\LJB for short).\\
A \LJB $(\L,\circ,[\cdot,\cdot]_\lambda)$ is in particular a JB--algebra (a Jordan--Banach algebra) when considered with the Jordan product alone. 

If $\L$ is a  LJB--algebra,  by taking combinations of the two products we can define an associative product on the complexification $\L^\mathbb{C}$.  
Specifically, we define:
$$ a b = a\circ b - i \sqrt{\kappa} [a,b]_\lambda.$$
Associativity follows from the Leibniz property (\ref{leibniz}) and the Jacobi identity (\ref{jacobi}).   Such associative algebra equipped with the norm $|| a + ib ||^2 = ||a ||^2 + || b ||^2$ becomes a $\CA$ whose real part is precisely $\L$.
When we consider a Lie--Jordan algebra to consist of all smooth functions on a classical carrier space, equipped with
the pointwise product $f\circ g\ (x) = f(x)g(x)$, and a Lie bracket of the form $[f,g] = \{ f, g \}$, with $\{ \cdot, \cdot \}$ being a Poisson 
bracket, then all previous requirements are satisfied with $\kappa = 0$.   It follows that from an algebraic point of view it is quite appropriate to consider a Poisson algebra as a Lie--Jordan algebra.\\

The space of states $\mathcal{S}(\L)$ of a Jordan algebra consists of all those real normalized positive linear functionals on $\L$, i.e. $$\rho: \L \longrightarrow \mathbb{R}$$ such that $\rho(\mathds{1}) = 1$ and $\rho(a^2) \geq 0 \ \forall a \in \L$. The state space of a JB--algebra is convex and compact with respect to the $\mathrm{w}^*$--topology. It was an important problem to determine which state spaces of Jordan algebras are the state spaces of a $\CA$. In fact it follows from results of Kadison \cite{Ka51} that if $\mathcal S(\A)$ denotes the state space of a $\CA$ $\A$, then $\A_{\mathrm{sa}}$ is naturally isomorphic to the space of affine $\mathrm{w}^*$--continous real functions on $\mathcal S(\A)$, hence the state space $\mathcal S(\A)$ as a topological space is determined just by the self-adjoint part of $\A$. The previous problem was solved by Alfsen and Shultz \cite{Al98} and A. Connes \cite{Co74} independently.  We will describe the reconstruction of the $\CA$ in the following section.

\section{Dynamical correspondence and \LJBs}

Following \cite{Al98} we will define a derivation of a JB--algebra $\L$ by focusing first only on the order structure, ignoring the algebraic multiplicative aspect. All the proofs contained in \cite{Al98} will be omitted. First note that a unital JB--algebra $\L$ is a complete order unit space with respect to the positive cone 
\begin{equation}\label{positive cone}
 \L^+ = \{a^2\ |\ a\in \L\}.
\end{equation}
\begin{definition}\label{order der}
A bounded linear operator $\delta$ on a JB--algebra $\L$ is called an order derivation if $e^{t\delta}(\L^+) \subset \L^+\ \forall t \in \mathbb{R}$.
\end{definition}

We denote the Jordan multiplier determined by an element $b \in \L$ by $\delta_b$. Thus for all $a \in \L$
$$\delta_b(a) = b \circ a.$$
It can be shown that Jordan multipliers $\delta_b$ are order derivations $\forall\ b \in \L$.

\begin{definition}
An order derivation $\delta$ on a unital JB--algebra $\L$ is self-adjoint if there exists $a \in \L$ such that $\delta = \delta_a$ and is skew-adjoint if $\delta(\mathds{1})= 0$.
\end{definition}

Again, it can be shown that if  $\delta$ is an order derivation, then $\delta$ is skew if and only if $\delta$ is a Jordan derivation, i.e., is a derivation with respect to the Jordan product, that is 
\begin{equation}\label{Jordander}
\delta(a \circ b) = \delta a \circ b + a \circ \delta b,  \quad \forall\ a,b \in \L .
\end{equation}

\begin{definition}
A dynamical correspondence on a unital JB--algebra $\L$ is a linear map
\begin{equation}
\psi: a \longrightarrow \psi_a 
\end{equation}
from $\L$ into the set of skew order derivations on $\L$ which satisfies
\begin{enumerate}[i)]
	\item There exists $k\in \R$ such that $k[\psi_a,\psi_b] = -[\delta_a,\delta_b] \ \forall\ a,b \in \L$, and,
\item $\psi_a a = 0 \ \forall\ a \in \L$.
\end{enumerate}
\end{definition}

It follows immediately from the definitions that:  
\begin{equation}\label{dynanti}
\psi_a b = - \psi_b a\ \forall\ a,b \in \L .
\end{equation}

\begin{definition}
Let $\L$ be a unital JB--algebra. A $C^*$--product compatible with $\L$ is an associative product on the complex linear space $\L \oplus i \L$ which induces the given Jordan product on $\L$ and makes $\L \oplus i \L$ into a $\CA$ with involution $(a + i b)^* = a - i b$ and whose norm induces the given norm in the JB-algebra $\mathcal{L}$.
\end{definition}

Note that if a JB--algebra $\L$ is the self-adjoint part of a $\CA$ $\A$, then there is a natural product and a norm induced in $\L \oplus i \L$ by using the representation $A = a + ib$ with $A \in \A,\ a,b \in \L$.  Such product and norm organizes $\L \oplus i \L$ into a $\CA$.  It follows that a JB--algebra is the self-adjoint part of a $\CA$ if and only if there exists a $C^*$--product compatible with $\L$ on $\L \oplus i \L$.    We can now state the main result in \cite{Al98}, relating the existence of a dynamical correspondence on a JB--algebra to the existence of a compatible $C^*$--product.

\begin{theorem}\label{dynC}
A unital JB--algebra $\L$ is Jordan isomorphic to the self-adjoint part of a $\CA$ if and only if there exists a dynamical correspondence on $\L$. Each dynamical correspondence $\psi$ on $\L$ determines a unique associative $C^*$--product compatible with $\L$ defined as
\begin{equation}\label{C*product}
ab = a\circ b- i\sqrt{k}\ \psi_a b
\end{equation}
and each $C^*$--product compatible with $\L$ arises in this way from a unique dynamical correspondence $\psi$ on $\L$.
\end{theorem}

We will now show that the existence of a dynamical correspondence on $\L$ is equivalent to the existence of a Lie product organizing $\L$ into a \LJB. 
First we need the following Lemmas:
\begin{lemma}
 Let $(\L,[\cdot,\cdot]_\L,\circ)$ be a \LJB. Then there exists an associative bilinear product on $\L \times \L$ defined as
\begin{equation}
 a \cdot b = a \circ b -i\sqrt{k} [a,b]_\L
\end{equation}
$\forall\ a,b \in \L$.
\end{lemma}

\medskip

{\parindent 0cm \emph{Proof:}}
 Bilinearity of the product follows directly from the bilinearity of the Jordan and Lie products. We have to prove the associativity, i.e.:
\begin{equation}
 a\cdot (b\cdot c) = (a\cdot b) \cdot c, \quad \forall\ a,b,c \in \L .
\end{equation}
The l.h.s. of the previous equation leads to:
\begin{equation*}
  a\cdot (b\cdot c) = a \circ (b \circ c) -i\sqrt{k}\ a \circ [b,c]_\L -i\sqrt{k}\ [a,b]_\L\circ c -i\sqrt{k}\ b \circ[a,c]_\L - k\ [a,[b,c]_\L]_\L,
\end{equation*}
and the r.h.s.:
\begin{equation*}
  (a\cdot b) \cdot c = (a \circ b) \circ c -i\sqrt{k}\ b\circ[a,c]_\L -i\sqrt{k}\ a \circ[b,c]_\L -i\sqrt{k}\ [a,b]_\L\circ c -k\ [[a,b]_\L,c]_\L,
\end{equation*}
Then
\begin{eqnarray*}
 a\cdot (b\cdot c) - (a\cdot b) \cdot c &=& a \circ (b \circ c) - (a \circ b) \circ c - k\ [a,[b,c]_\L]_\L - k\ [c,[a,b]_\L]_\L\\
&=& [b,[c,a]_\L]_\L + [a,[b,c]_\L]_\L + [c,[a,b]_\L]_\L\\
&=& 0.
\end{eqnarray*}
where we have used (\ref{jacobi}),(\ref{leibniz}) and (\ref{associator}).
\hfill$\Box$ \bigskip

 Note that the Jordan and Lie products can be also expressed in terms of the associative product as:
\begin{equation}\label{Jass}
 a \circ b = \frac{1}{2}(a\cdot b + b\cdot a),
\end{equation}
\begin{equation}\label{Lass}
 [a, b]_\L = \frac{i}{2\sqrt{k}}(a\cdot b - b\cdot a).
\end{equation}

\begin{lemma}\label{Jauto}
 Let $(\L,[\cdot,\cdot]_\L,\circ)$ be a \LJB. Then $e^{[a,\cdot]_\L}$ is a Jordan automorphism $\forall\ a \in \L$.
\end{lemma}

\medskip

{\parindent 0cm \emph{Proof:}}
 We have to prove that
\begin{equation}\label{automorphism}
 e^{[a,\cdot]_\L}\ (b\circ c) = (e^{[a,\cdot]_\L}\ b)\circ(e^{[a,\cdot]_\L}\ c).
\end{equation}
By Hadamard's formula, the l.h.s. of the previous equation is:
\begin{equation*}
e^{[a,\cdot]_\L}\ (b\circ c) = e^a\ (b \circ c)\ e^{-a}.
\end{equation*}
By using formula (\ref{Jass}), the r.h.s. of  (\ref{automorphism})  becomes:
\begin{eqnarray*}
 (e^{[a,\cdot]_\L}\ b)\circ(e^{[a,\cdot]_\L}\ c) &=& (e^a\ b\ e^{-a}) \circ (e^a\ c\ e^{-a})\\
&=& \frac{1}{2}\ e^a\ (b\cdot c)\ e^{-a} + \frac{1}{2}\ e^a\ (c\cdot b)\ e^{-a}\\
&=& e^a\ (b \circ c)\ e^{-a}.
\end{eqnarray*}
\hfill$\Box$ \bigskip

\begin{lemma}
Let $(\L,[\cdot,\cdot]_\L,\circ)$ be a \LJB. Then $[a,\cdot]_\L$ is an order derivation on $\L\ \forall\ a \in \L$.
\end{lemma}

\medskip

{\parindent 0cm \emph{Proof:}}
From Definition \ref{order der}, we have to prove that $e^{t[a,\cdot]_\L}\ (\L^+) \subset \L^+$ $\forall\ a\in \L$ and $\forall\ t \in \R$. Since $e^{t[a,\cdot]_\L}$ is a Jordan automorphism (Lemma \ref{Jauto}), we have:
\begin{equation*}
 e^{t[a,\cdot]_\L} (b \circ b) = (e^{t[a,\cdot]_\L} b) \circ (e^{t[a,\cdot]_\L} b)
\end{equation*}
$\forall\ a,b \in \L$ and $\forall\ t\in\R$, i.e. $e^{t[a,\cdot]_\L}$ preserves the positive cone (\ref{positive cone}) $\L^+$.
\hfill$\Box$ \bigskip

\begin{theorem}\label{LJBdyn}
Let $\L$ be a unital JB--algebra. There exists a dynamical correspondence $\psi$ on $\L$ if and only if $\L$ is a {\LJB}  with Lie product $[\cdot,\cdot]_\L$ such that 
\begin{equation}\label{lie_dynamical}
[a,b]_\L= \psi_a b
\end{equation}
\end{theorem}

\medskip

{\parindent 0cm \emph{Proof:}}
First assume that $\L$ is a \LJB. From Definition \ref{dynC} we have to check that $\forall\ a,b \in \L$
$$[\psi_a,\psi_b] = -[\delta_a,\delta_b]$$
that is
$$[a,[b,c]_\L]_\L - [b,[a,c]_\L]_\L = b \circ (a \circ c) - a \circ (b \circ c)$$
which is an easy computation once the Jordan and Lie products are expressed as in (\ref{Lass}) and (\ref{Jass}). From the antisymmetry of the Lie product it is also true that $\psi_a a = [a,a]_\L = 0\ \forall\ a\in \L$. Hence the linear map $a \longrightarrow [a,\cdot]_\L$ from the {\LJB} $\L$ to the skew-order derivations on $\L$ is a dynamical correspondence.\\
Conversely, assume $\L$ is a JB--algebra with a dynamical correspondence $\psi$. Then from (\ref{dynanti}) $\psi_a b = [a,b]_\L$ is antisymmetric. By using the expression \ref{C*product} it is an easy computation to check that the Jacobi property (\ref{jacobi}) is satisfied
$$\psi_a(\psi_b c) + \psi_b(\psi_c a) + \psi_c(\psi_a b) = 0.$$
The Leibniz identity (\ref{leibniz}) follows from (\ref{Jordander}) and also the compatibility condition (\ref{associator}) is easy to check with a simple computation. Hence a JB--algebra with a dynamical correspondence is a \LJB.
\hfill$\Box$ \bigskip

\begin{corollary}\label{isoC}
 A unital JB--algebra $\L$ is Jordan isomorphic to the self-adjoint part of a $\CA$ if and only if it is a \LJB.
\end{corollary}
\medskip

{\parindent 0cm \emph{Proof:}}
 This is an obvious consequence of Theorems \ref{dynC} and \ref{LJBdyn}.
\hfill$\Box$ \bigskip

\begin{corollary}\label{complexification}
 Let $(\L,\circ,[\cdot,\cdot]_\L)$ be a {\LJB} and $\mathcal A = \L^\mathbb{C}$ the natural $\CA$ defined by the complexification of $\L$. Then there is a natural identification between the states $\mathcal S(\L)$ of $\L$ and the states $S(\mathcal A)$ of the  $\CA$ $\mathcal A$.
\end{corollary}
\medskip

{\parindent 0cm \emph{Proof:}}
  Given a state $\rho$ of $\L$, we define a linear functional $\tilde{\rho}$ of $\mathcal A$ by extending it linearly.   The linear functional $\tilde{\rho}$ is positive and normalized because $\rho$ is positive and normalized.  The converse is trivial.
 
\hfill$\Box$ \bigskip

\section{Reduction of Lie--Jordan algebras by Jordan ideals}\label{LJ red}

The algebraic reduction of a Lie--Jordan algebra should start from the consideration of both products, the Jordan $\circ$ and the Lie product $[\cdot, \cdot ]_\L$.  As usual to reduce a Jordan
algebra $\L$ we have to identify a Jordan ideal, say $\J$, and then form the quotient algebra
$$  \L/ \J $$
that carries the structure of a Jordan algebra, however such space will not inherit a Lie bracket in general.   The space of all Jordan derivations of $\L$ carries a natural structure of Lie algebra.  We consider the subalgebra of those derivations which map $\J$ into itself.  By using the Lie product, i.e., the dynamical correspondece, it is possible to associate with any element $a$ in $\L$ a Jordan derivation\footnote{In the following we will denote the Lie brackets $[\cdot,\cdot]_\L$ simply by $[\cdot,\cdot]$ without ambiguity.} as in eq. (\ref{lie_dynamical}),
$  \psi_a(b) =   [a, b ]$.   This association is a Lie algebra homomorphism, in fact: 
$$ \psi_{[a,b]} = [\psi_a, \psi_b].$$
Define the normalizer of $\J$ as
\begin{equation}
\N = \{a \in \L\ |\ [a, \J] \subset \J\}.
\end{equation}

\begin{lemma}
The normalizer $\N$ of a Jordan ideal $\J$ of a \LJB $\L$ is a closed unital Lie-Jordan subalgebra of $\L$.
\end{lemma}

\medskip

{\parindent 0cm \emph{Proof:}}
The normalizer $\N$ is a Lie subalgebra of $(\L, [\cdot, \cdot ]_{\L})$.  In fact due to the Jacobi identity (\ref{jacobi}), $[\N,\N] \subset \N$:
$$[[a,b],\J] \subset [a,[b,\J]] + [b,[\J,a]] \subset [a,\J] + [b,\J] \subset \J \quad \forall a,b \in \N .$$
The normalizer is also a Jordan subalgebra, in fact due to the Leibniz identity (\ref{leibniz}), $\N \circ \N \subset \N$:
$$[a \circ b,\J] \subset a \circ [b,\J] + b \circ [a,\J] \subset a \circ \J + b \circ \J \subset \J \quad \forall a,b \in \N .$$
It is also a Lie--Jordan subalgebra since it verifies the compatibility conditions (\ref{leibniz}) by definition of Jordan derivation and (\ref{associator}) because it is a subalgebra of $\L$. It also contains the Jordan unit since it is annihilated by the Lie product. Finally, because the Lie product is continuous, we have that the normalizer $\N$ is closed.
\hfill$\Box$ \bigskip

It is now possible to consider those derivations associated with $\J$ which preserve $\J$, i.e., $\t{\J} = \N \cap \J$ is the subset of $\J$ such that:
$$ [\t{\J}, \J] \subset \J. $$

\begin{lemma}  Let $\J$ be a closed Jordan ideal of the \LJB  $\L$.  Then
the subspace $\t{\J} = \N \cap \J$ is a closed Lie-Jordan ideal of the \LJB $\N$.
\end{lemma}

\medskip

{\parindent 0cm \emph{Proof:}}   The subspace $\t{\J} = \N \cap \J$ is closed because both $\N$ and $\J$ are closed sets.
Because $\J$ is a Jordan ideal, given $a \inÊ\N$, we will have $a\circ \N \cap \J \subset \J$ and because $\N$ is a Jordan subalgebra, we will get $a\circ \N \cap \J \subset \N$, hence $a\circ\t{\J} \subset \t{\J}$ for all $a \in \N$.  Similarly, it is shown that $\t{\J}$ is a Lie algebra ideal of $\N$.
\hfill$\Box$ \bigskip

The closed Lie-Jordan ideal $\t{\J}$ of the \LJB $\N$ induces a canonical Lie--Jordan algebra structure in the quotient:
\begin{equation}
\t\L = \bigslant{\N}{\N \cap \J}.
\end{equation}
We will denote elements of $\t\L$ by $\t a$.\\
Observe that all the reduction procedure works also for the case in which $\L$ is a \mbox{(pre-) Lie--Jordan} algebra (see section \ref{second}) since we did not make use of the compatibility condition (\ref{associator}) and the reduced algebra $\t\L$ will be a (pre-) Lie--Jordan algebra as well.\\
The quotient Lie-Jordan algebra $\t\L$ carries the quotient norm, 
$$\|\t a\| = \| [a]\| = \displaystyle\inf\limits_{b \in \N \cap \J} \|a +b \|  ,$$ 
where $a \in \N$ is an element of the equivalence class $[a]$ of $\N$ with respect to the ideal $\t\J = \N \cap \J$.
The quotient norm provides a \LJB structure to $\t\L$.

Hence the reduction of the Lie--Jordan algebra $\L$ with respect to the closed Jordan ideal $\J$ is given by the short exact sequence
\begin{equation}
0 \to \t{\J} \to \N  \to \t \L \to 0 
\end{equation}
and $\t\L$ is the reduction algebra of $\L$, that will be denoted as $\L /\hskip-.07cm / \J$ too.\\

\section{Reduction of Lie--Jordan algebras by Lie--Jordan subalgebras}\label{redsub}

The reduction procedure of the previous section is given by means of a Jordan ideal $\J$ of the given Lie--Jordan algebra $\L$. Actually the assumption for $\J$ being an ideal is not necessary to obtain a reduced Lie--Jordan algebra. We will prove in this section that it is still possible to obtain a reduction of $\L$ by relaxing the ideal property of $\J$ and requiring it to be a Lie--Jordan subalgebra of $\L$.

\begin{theorem}\label{JL sub}
Let $\L$ be a Lie--Jordan algebra and $\J$ a non-unital Lie--Jordan subalgebra of $\L$. Then $\N \circ \J$ is a Lie--Jordan ideal of $\N$.
\end{theorem}

\medskip

{\parindent 0cm \emph{Proof:}}
The normalizer $\N$ is a unital Lie--Jordan subalgebra due to Jacobi (\ref{jacobi}) and Leibniz (\ref{leibniz}) identities. Then it follows that $\J$ is a proper subspace of $\N$, since it is obtained by multiplication with the Jordan unit $\mathds{1} \in \N$: $\J = \mathds{1} \circ \J$. 

The subspace $\N \circ \J$ is a Lie ideal of $\N$, i.e.
$$[\N \circ \J, \N] \subset \N \circ \J.$$
In fact by using the Leibniz property (\ref{leibniz}): $$[\N \circ \J, \N] \subset [\N, \N] \circ \J + [\J, \N] \circ \N \subset \N \circ \J.$$
The subspace $\N \circ \J$ is also a Jordan ideal of $\N$, in fact by using the compatibility condition (\ref{associator}):
$$(\N \circ \J) \circ \N \subset \J \circ (\N \circ \N) + [\N, [\N, \J]] \subset \J \circ \N + \J,$$
and since $\J \subset \N$ we obtain $$(\N \circ \J) \circ \N \subset \N \circ \J.$$
Note that $\N \circ \J$ also satisfies the compatibility conditions (\ref{leibniz}) and (\ref{associator}) and hence it is a Lie--Jordan subalgebra.
\hfill$\Box$ \bigskip

Because of the previous Theorem, it is canonically defined a Lie--Jordan algebra structure in the quotient space
$$\t \L = \bigslant{\N}{\N \circ \J} ,$$
and it is an unital algebra. As in the previous section if $\L$ is a \LJB and $\J$ is a closed subspace, then $\t\L$ inherits a \LJB algebra structure with respect to the quotient norm.\\
Observe that if $\J$ is a unital Lie--Jordan subalgebra, then $\N \circ \J$ would be a trivial ideal of $\N$ since they would coincide and the reduction would have no sense.

\section{Reduction of Lie--Jordan algebas and constraints in $\CA$s}\label{constraints}

Following \cite{Grundling:1984sq}, \cite{Grundling:1998zn} we briefly recall how to deal with quantum constraints in a $\CA$ setting.  The aim of this section is to prove that the reduction procedure of $\CA$s used to analize quantum constraints, also called T--reduction, can be equivalently described by using the theory of reduction of \LJBs discussed above.

A quantum system with constraints is a pair $(\F,\mathcal C)$ where the field algebra $\F$ is a unital $\CA$ containing the self--adjoint constraint set 
$\mathcal C$, i.e. $C=C^*\ \forall C\in{\mathcal C}$.
The constraints select the physical state space, also called Dirac states
$$ \S_D \equiv \{\omega \in \S(\F)\ |\ \omega(C^2) = 0\ \ \forall C \in \mathcal C \}$$
where $\S(\F)$ is the state space of $\F$.

Define $\D = \left[\F \mathcal C\right] \cap \left[\mathcal C \F\right]$ where the notation $\left[\ \cdot\ \right]$ denotes the closed linear space generated by its argument.  Then we have:

\begin{theorem}
 $\D$ is the largest non-unital $\CA$ in $\displaystyle\bigcap \limits_{\omega \in \S_D} \emph{ker}\ \omega$.
\end{theorem}

For any set $\Omega \in \F$, define as before its normalized or ``weak commutant'' as
\begin{equation}
\Omega_W = \{F \in \F\ | \ [F,H]  \in  \Omega\ \ \forall H\in \Omega\}
\end{equation}
which corresponds roughly to Dirac's concept of ``first class variables'' \cite{Dirac}.
Consider now the multiplier algebra of $\Omega$ as
\begin{equation}
\M(\Omega) = \{ F \in \F\ |\ FH\in \Omega,\ HF \in \Omega\ \ \forall H\in \Omega\} 
\end{equation}
i.e. the largest set for which $\Omega$ is a bilateral ideal. $\M(\Omega)$ is clearly an unital $\CA$ and we obtain:
\begin{theorem}\label{DW=MD}
 $\O \equiv \D_W = \M(\D)$
\end{theorem}
That is, the weak commutant of $\D$ is also the largest set for which $\D$ is a bilateral ideal and it will be denoted by $\O$.   It follows that the maximal (and unital) $\CA$ of physical observables determined by the constraints $\mathcal{C}$ is given by:
\begin{equation}
 \t\F = \O / \D.
\end{equation}

We will now show that this procedure is equivalent to a \LJB reduction by a Lie--Jordan subalgebra as discussed in section \ref{redsub}).

Define $\L$ and $\t\L$ such that $\F = \L \oplus i \L$ and $\t\F = \t\L \oplus i\t\L$, i.e. they are the self-adjoint part of $\F$ and $\t\F$ respectively. From Corollary \ref{isoC} it follows that $\L$ and $\t\L$ are unital \LJBs.  
Similarly define the \LJBs $\N$ and $\J$ as the self-adjoint parts of $\O$ and $\D$ respectively, i.e. $\O = \N \oplus i\N$, $\D = \J \oplus i\J$.\\

\begin{lemma}\label{ODNJ}  Let $\N$ and $\J$ be two Lie-Jordan subalgebras of $\L$.  Then
 $\O = \N \oplus i\N$ is the weak commutant of $\J \oplus i\J$ if and only if $\N = \J_W$.
\end{lemma}
\medskip

{\parindent 0cm \emph{Proof:}}
First let us remark that the normalizer is a vector space. Let $a +ib \in \O$ with $a,b \in \N$. By definition
$$[a+ib,\J \oplus i \J] \subset \J \oplus i\J$$
that is $$[a+b,\J] \subset \J \quad \mbox{and} \quad [a-b,\J]\subset \J.$$
This implies $$[a,\J] \subset \J \quad \mbox{and}\quad [b,\J]\subset\J \quad \forall\ a,b\in \N$$ that is the self-adjoint part of $\O$ is the normalizer of the self-adjoint part of $\D$. Conversely $$[a,\J] \subset \J \quad \forall\ a \in \N$$
implies that 
$$[a+b,\J] \subset \J \quad \mbox{and} \quad [a-b,\J]\subset \J\quad \forall\ b\in \N$$
i.e.
$$[a+ib,\J \oplus i\J] \subset \J \oplus i \J.$$
\hfill$\Box$ \bigskip

\begin{lemma}\label{ideals}
 Let $\N$ and $\J$ be two Lie-Jordan subalgebras of $\L$. Then $\J$ is a Lie--Jordan ideal of $\N$ if and only if $\D = \J \oplus i \J$ is an associative bilateral ideal of $\O = \N \oplus i\N$.
\end{lemma}
\medskip

{\parindent 0cm \emph{Proof:}}
Using the expressions provided by eqs. (\ref{Jass}) and (\ref{Lass}), the statement becomes an easy computation.
\hfill$\Box$ \bigskip

\begin{theorem}  With the notations above, let $\F$ be a $\CA$ and $\mathcal{C}$ a real constraint set.   Let $\O = \D_W$ and $\D = [\F \mathcal{C}] \cap [\mathcal{C}\F]$ be as in Thm. \ref{DW=MD}.   Let as before $\t\L$ denote the real part of the $\CA$  $\O/\D$.   Then:
 $$\t\L = \N / \J ,$$
 with $\O = \N \oplus i\N$ and $\D = \J \oplus i\J$.
\end{theorem}

\medskip

{\parindent 0cm \emph{Proof:}}
 $\J$ is a LJB--subalgebra of $\L$ and hence from the previous Lemma \ref{ideals} it is a Lie--Jordan ideal of $\N$. From the Lemma \ref{ODNJ}, it also follows that $\N$ is its normalizer. Hence the reduced space $\N / \J$ is a \LJB whose complexification is $\t\F$ since
$$ \N / \J \oplus i\ \N / \J = \bigslant{\N \oplus i\ \N}{\J \oplus i\ \J} = \O / \D.$$
It follows form Cor. \ref{isoC} that $\t\L = \N / \J$.
\hfill$\Box$ \bigskip

Thus it is evident that the $\CA$ reduction gives rise to a reduction of the \LJB by a Lie--Jordan subalgebra $\J$ and, conversely, a reduction of a \LJB with respect to a Lie--Jordan subalgebra $\J$ always gives rise to a $\CA$ reduction with respect to $\D$. The two approaches are therefore completely equivalent in this sense, as the following ``functorial'' diagram shows:\\
$$
 \xymatrix@C=70pt@R=60pt{
\L \ar@{-}[d]_{{\J\ \mathrm{LJ subalgebra\ }\\ \\ \ \N}} |-{\SelectTips{cm}{}\object@{>}} 
\ar@{-}[r] |-{\SelectTips{cm}{}\object@{<}}  & \F = \L \oplus i \L
 \ar@{-}[d]^{{\D\ \mathrm{ass. ideal}\\ \\ \ \O\ =\ \N\ \oplus\ i\ \N}} |-{\SelectTips{cm}{}\object@{>}} \\
\t\L \ar@{-}[r] |-{\SelectTips{cm}{}\object@{>}} & \t\F = \O / \D = \t\L \oplus i\t\L }
$$
\vspace{0.2cm}
s
Let us now consider the reduction of a \LJB by a Jordan ideal first and then pass to the complexified $\CA$. We will obtain in this way an ``extension'' of the $\CA$ reduction procedure. Let $\L$ be the initial \LJB with Jordan ideal $\J$, whose normalizer $\N$ enables us to obtain the reduced \LJB $\t\L = \bigslant{\N}{\N \cap \J}$.

Define now $\O = \N \oplus i \N$ and $\D = \J \oplus i \J$. From Lemma \ref{ODNJ}, $\O$ is the normalizer of $\D$ and from Lemma \ref{ideals},$\D \cap \O$ is an associative bilateral ideal of $\O$.

We obtain immediately the following:

\begin{theorem}\label{reduc_ideal}  With the notations above, given an ideal $\J$ of the \LJB $\L$, the reduced \LJB $\t\L = \L /\hskip -0.07cm/ \J$ defines a $\CA$ $\t\F$ that is obtained from the field algebra $\F = \L \oplus i \L$ as:
$$\t\F = \t\L \oplus i\t\L = \bigslant{\N \oplus i\ \N}{\N \cap \J \oplus i\ \N \cap \J} = \bigslant{\O}{\D \cap \O} .$$
\end{theorem}

Again, the following diagramme summarizes the discussion above:
$$
 \xymatrix@C=50pt@R=70pt{
\L \ar@{-}[d]_{{\J\ \mathrm{J. Ideal\ }\\ \\ \ \N}} |-{\SelectTips{cm}{}\object@{>}} 
\ar@{-}[r] |-{\SelectTips{cm}{}\object@{>}}  & \F = \L \oplus i \L
 \ar@{-}[d]^{{\D\ =\ \J\ \oplus\ i\ \J\\ \\ \ \O\ =\ \N\ \oplus\ i\ \N}} |-{\SelectTips{cm}{}\object@{>}} \\
\t\L=\bigslant{\N}{\N \cap \J} \ar@{-}[r] |-{\SelectTips{cm}{}\object@{<}} & \t\F = \t\L \oplus i\t\L = \bigslant{\D}{\D \cap \O}}
$$
\vspace{0.1cm}

Note that if in addition $\J$ is a Lie--subalgebra then $\D \subset \O$ and $\D$ would be a bilateral $C^*$--ideal, recovering the usual reduction procedure of $\CA$s.

\section{The space of states of a reduced \LJB}\label{reduced_states}

The purpose of the remaining two sections is to discuss the structure of the space of states and the GNS construction of reduced states for reduced \LJBs 
with respect to the space of states of the unreduced \LJB.

As it was discussed in the previous section, let $\A$ be a $\CA$, $\L=\A_{\mathrm{sa}}$ its real part and $\J$ a Jordan ideal of $\L$.  Let $\mathcal{N}_{\J}$ be the normalizer of $\J$ with respect to the Lie algebra structure of the Lie--Jordan algebra $(\L,\circ,[\cdot,\cdot]_\L)$. Then we will denote as before by $\t{\L}$ the reduced Lie--Jordan Banach algebra $\N /( \N \cap \J)$ and its elements by $\t a$.

Let $\t{\mathcal{S}} = \mathcal{S}(\t{\L})$ be the state space of the reduced \LJB $\t{\L}$, i.e. $\t{\omega} \in \t{\mathcal S}$ means that $\t{\omega}(\t a^2) \geq 0 \ \forall \t a \in \t{\L}$, and $\t{\omega}$ is normalized. Notice that if $\L$ is unital, then $\mathds{1} \in \N$ and $\mathds{1} + \N \cap \J$ is the unit element of $\t{\L}$. We will denote it by $\t{\mathcal{\mathds{1}}}$.\\ We have the following:

\begin{lemma}\label{lemma1}
 There is a one-to-one correspondence between normalized positive linear functionals on $\t{\L}$ and normalized positive linear functionals on $\N$ vanishing on $\N \cap \J$.
\end{lemma}

\medskip

{\parindent 0cm \emph{Proof:}}
 Let $\omega' : \N \longrightarrow \mathbb{R}$ be positive. 
The positive cone on $\t{\L}$ consists of elements of the form $\t a^2 = (a + \N \cap \J)^2 = a^2 + \N \cap \J$, i.e.$$\mathcal K^+_{\t{\L}} = \{a^2 + \N \cap \J\ |\ a \in \N \} = \mathcal K^+_{\N} + \N \cap \J.$$
Thus if $\omega'$ is positive on $\N$, $\omega'(a^2)\geq 0$, hence:$$\omega'(a^2 + \N \cap \J) = \omega'(a^2) + \omega'(\N \cap \J)$$ and if $\omega'$ vanishes on the closed ideal $\N \cap \J$, then $\omega'$ induces a positive linear functional on $\t{\L}$.  Clearly $\omega'$ is normalized then the induced functional is normalized too because $\t{\mathcal{\mathds{1}}} = \mathds{1} + \N \cap \J$. 

Conversely, if $\t{\omega}:\t{\L} \longrightarrow \mathbb{R}$ is positive and we define
$$\omega'(a) = \t{\omega}(a + \N \cap \J)$$
then $\omega'$ is well-defined, positive, normalized and $\omega'|_{\N \cap \J}=0$.
\hfill$\Box$ \bigskip

Now we will relate positive linear functionals on $\N$ vanishing on $\N \cap \J$ with a class of linear functionals on $\L$. For this purpose, we will first notice that given a positive linear functional on $\N$ there exists an extension of it to $\L$ which is positive too.

\begin{lemma}\label{lemma2}
 Given a closed Jordan subalgebra $\N$ of a JB--algebra $\L$ such that $\mathds{1} \in \N$ and $\omega' $ a normalized positive linear functional on $\N$, then there exists $\omega: \L \longrightarrow \mathbb R$ such that $\omega(A) = \omega'(A) \ \forall A \in \N$ and $\omega \geq 0$.
\end{lemma}
\medskip

{\parindent 0cm \emph{Proof:}}
 Since $\L$ is a JB--algebra, it is also a Banach space. Due to the Hanh--Banach extension theorem, there exists a continous extension $\omega$ of $\omega'$, i.e. $\omega(A) = \omega'(A) \ \forall A \in \N$, and moreover $\|\omega\|=\|\omega'\|$.

From the equality of norms and the fact that $\omega'$ is positive we have
$\|\omega\|=\omega'(\mathds{1})$,
but $\omega$ is an extension of $\omega'$ then $\|\omega\|=\omega(\mathds{1})$, 
which implies that $\omega$ is a positive functional and satisfies all the requirements
stated in the lemma.
\hfill$\Box$ \bigskip

We will denote by $\mathcal S((\N\cap \J)^0)$ the set of states vanishing on $\N\cap \J$.   We will consider the equivalence classes of states on $\mathcal S((\N\cap \J)^0)$ (which is again a convex set) with respect to the equivalence relation induced by $\N^0$.  This is, the canonical projection
$$\pi \colon (\N\cap \J)^0 \longrightarrow \bigslant{(\N\cap \J)^0}{\mathcal N^0_{\J}}$$
 induces an equivalence relation on $\mathcal S(( \N\cap \J)^0)$ whose equivalence classes are $\pi^{-1}([\omega])\cap\mathcal S(\L)$ with $[\omega] \in (\N\cap \J)^0/\mathcal N^0_{\J}$. We will call this equivalence relation the $\N^0$--equivalence relation on $\mathcal S(( \N\cap \J)^0)$.  Hence we have proved the following:
 
\begin{theorem}
 There is a one-to-one correspondence between the states $\t\omega$ of the reduced Lie--Jordan Banach algebra $\t{\L}$ and the $\mathcal N^0_{\J}$--equivalence classes of states $\omega \in \mathcal S((\N\cap \J)^0) \subset \mathcal S(\L)$.
\end{theorem}

\medskip

{\parindent 0cm \emph{Proof:}}
  The proof is almost trivial using the results above.  Thus if $\omega$ is an state on $\mathcal S(( \N\cap \J)^0)$, then its restriction to $\N$ vanishes on $\N\cap \J$ and because of Lemma  \ref{lemma1} it induces an state on $\t\L$.

Conversely, states on $\t\L$ are in one--to--one correspondence with states on $\N$ vanishing on $\N\cap \L$, hence because of Lemma \ref{lemma2} there exists an extension of any such state to a state on $\L$.  However two such extension define the same state on $\N$ if they differ on an element of $\N^0$, hence they are $\N^0$--equivalent.
\hfill$\Box$ \bigskip

\section{The GNS representation of reduced states}\label{reduced_gns}

Finally, we will try to describe the GNS representation of a reduced state in terms of data from the unreduced \LJB. Let $\t \L$ be, as before, the reduced LJB--algebra of $\L$ with respect to the Jordan ideal $\J$.  Denote by $\t \A = \t \L \oplus i\t \L$ the corresponding $\CA$ and by $\t\S$ its state space. Let $\t \omega \in \t \S$ be a normalized state on $\t \A$. The GNS representation of $\t \A$ associated to the state $\t\omega$, denoted by 
$$ \pi_{\t \omega}: \t \A \longrightarrow \mathcal{B}(\Hw),$$
is defined as 
$$ \pi_{\t \omega}(\t A)(\t B + \J_{\t \omega}) = \t A\t B + \J_{\t \omega} \quad \forall\t A,\t B \in \t\A,$$
where the Hilbert space $\mathcal H_{\t \omega}$ is the completion of the pre--Hilbert space defined on $\t\A / \J_{\t \omega}$ by the inner product
$$\langle \t A + \J_{\t \omega},\t B + \J_{\t \omega}\rangle \equiv \t\omega(\t A^*\t B)$$
and $\J_{\t\omega}$ is the Gelfand ideal of $\t\omega$. Let $\omega$ be a state on $\A$ that extends the state $\omega'$ on $\Nc$ induced by $\t\omega$ according to Lemmas \ref{lemma1} and \ref{lemma2}. Notice that $\omega|_{\N}$ vanishes on $\N \cap \J$, thus the Gelfand ideal $\J_\omega$ of $\omega$ contains $\N \cap \J$. We will have then:
\begin{theorem}
 There is a unitary equivalence between $\Hw$ and the completion of the pre--Hilbert space:
$$\bigslant{\mathcal H' = (\N + \J)^{\C}}{(\N + \J)^{\C} \cap \J_{\omega}}$$
with the inner product defined by
$$\langle A + \Nc \cap \J_{\omega},B + \Nc \cap \J_{\omega}\rangle' \equiv \t\omega(A^* B)$$
$\forall A,B \in (\N + \J)^{\C}$.
\end{theorem}
\medskip

{\parindent 0cm \emph{Proof:}}
 Notice first that $\langle \cdot, \cdot \rangle'$ is well defined because of the properties of the Gelfand ideal $\J_{\omega}$. Morover we have that
$$\Hw = \t\A / \J_{\t \omega}$$
but because of Thm. \ref{reduc_ideal}, $\t A = \Nc/(\Nc \cap \J^\C)$ and $\J_{\t\omega} = \J_{\omega'}/(\J_{\omega'}\cap\J^{\C})$.\\
Hence because $\J_{\omega'}=\J_{\omega}\cap\Nc$ and $\J^\C \subset \J_{\omega}$, we have:
\begin{eqnarray*}
 \Hw &=& \t\A / \J_{\t \omega} = \bigslant{\bigslant{\Nc}{\Nc \cap \J^\C}}{\bigslant{\J_{\omega}\cap\Nc}{\J_{\omega}\cap\Nc\cap\J^{\C}}}\\
     &\cong& \bigslant{\Nc + \J^\C/\J^\C}{\J_{\omega}\cap\Nc + \J^\C/\J^\C} \\
     &\cong& \bigslant{\Nc + \J^\C}{\J_{\omega}\cap\Nc + \J^\C \cap \J_{\omega}} = \bigslant{(\N + \J)^\C}{(\N + \J)^\C \cap \J_{\omega}}.
\end{eqnarray*}
\hfill$\Box$ \bigskip

Notice that
\begin{eqnarray*}
 \mathcal H' &=& \bigslant{(\N + \J)^{\C}}{(\N + \J)^{\C} \cap \J_{\omega}} \cong \bigslant{(\N + \J)^\C + \J_{\omega}}{\J_{\omega}}\\ &\cong& \bigslant{\Nc + \J^\C + \J_{\omega}}{\J_{\omega}}.
\end{eqnarray*}
Thus the reduced GNS construction corresponding to the state $\t\omega$ is the GNS construction of any extension $\omega$ of $\t\omega$ restricted to $\Nc + \J^\C + \J_\omega$. Notice that $\t\omega$ will be a pure state if and only if $\pi_{\t\omega}$ is irreducible, i.e. if the representation of $\pi_{\omega}$ of $\omega$ restricted to $\Nc + \J^\C + \J_\omega$ is irreducible. Then if $\Nc + \J^\C + \J_\omega = \A$, $\pi_\omega$ will be irreducible if $\omega$ is a pure state. If $\Nc + \J^\C + \J_\omega \varsubsetneq \A$, then the state $\omega$ extending $\t\omega$ cannot be pure.

\vskip 0.5cm

\paragraph{\bf Acknowledgements:} 
The authors would like to thank F. Lled\'o by various valuable comments that have helped to prepare the final version of this paper.
This work was partially supported by MEC grant
MTM2010-21186-C02-02 and QUITEMAD programme.  G.M. would like to acknowledge the support provided by the Santander/UCIIIM Chair of Excellence programme 2011-2012.

\end{document}